\documentclass[letterpaper, 10 pt, conference]{ieeeconf}
\IEEEoverridecommandlockouts
\usepackage{graphics} 
\usepackage{epsfig} 
\usepackage{mathptmx} 
\usepackage{amsmath} 
\usepackage{amssymb}  
\usepackage{multirow}
\usepackage{bbding}
\usepackage{graphicx}
\usepackage{subcaption}
\usepackage{flushend}
\usepackage{hyperref}
\usepackage{xcolor}
\usepackage{cite}
\usepackage{jabbrv}
\hypersetup{
    colorlinks,
    linkcolor={red!50!black},
    citecolor={blue!50!black},
    urlcolor={blue!80!black}
}

\title{\LARGE \bf
A cognitive process approach to modeling gap acceptance in overtaking
}

\author{Samir H.A. Mohammad$^{1}$, Haneen Farah$^{2}$ and Arkady Zgonnikov$^{1}$
\thanks{$^{1}$Samir H.A. Mohammad and Arkady Zgonnikov are with the Department of Cognitive Robotics, Faculty of Mechanical, Maritime and Materials Engineering, Delft University of Technology, Mekelweg 2, Delft, 2628CD, the Netherlands.
        {\tt\small a.zgonnikov@tudelft.nl}}%
\thanks{$^{2}$Haneen Farah is with the Department of Transport and Planning, Faculty of Civil Engineering and Geosciences, Delft University of Technology, Stevinweg 1, Delft, 2628CN, the Netherlands.}
}

\begin{document}

\maketitle
\thispagestyle{empty}
\pagestyle{empty}

\begin{abstract}

Driving automation holds significant potential for enhancing traffic safety. However, effectively handling interactions with human drivers in mixed traffic remains a challenging task. Several models exist that attempt to capture human behavior in traffic interactions, often focusing on gap acceptance. However, it is not clear how models of an individual driver's gap acceptance can be translated to dynamic human-AV interactions in the context of high-speed scenarios like overtaking. In this study, we address this issue by employing a cognitive process approach to describe the dynamic interactions by the oncoming vehicle during overtaking maneuvers. Our findings reveal that by incorporating an initial decision-making bias dependent on the initial velocity into existing drift-diffusion models, we can accurately describe the qualitative patterns of overtaking gap acceptance observed previously. Our results demonstrate the potential of the cognitive process approach in modeling human overtaking behavior when the oncoming vehicle is an AV. To this end, this study contributes to the development of effective strategies for ensuring safe and efficient overtaking interactions between human drivers and AVs.

\end{abstract}

\section{Introduction}
Driving automation has the potential to enhance traffic safety \cite{MILAKIS2017324}. However, as the road will continue to consist of mixed traffic in the foreseeable future, effectively handling interactions between automated vehicles (AVs) and human drivers remains a significant challenge. Understanding how humans behave in these interactions is crucial to address this problem, particularly in high-stakes scenarios such as overtaking maneuvers \cite{Schieben2019}.

Many models of human behavior in traffic interactions have been proposed, with a particular focus on gap acceptance as a key aspect of the interaction (e.g., \cite{Farah2010, Stefansson2020}). These models have provided valuable insights into human driver behavior. However, the translation of individual driver gap acceptance models to human-AV interactions remains unclear. While recent research has started addressing this issue (e.g., \cite{Markkula2020, nudge}, it has primarily focused on low-speed interactions, limiting its applicability to high-speed scenarios like overtaking.

Therefore, there is a need to gain a better understanding of the interactions between oncoming AVs and human drivers during overtaking maneuvers. In our study, through a conceptual analysis of interactions during overtaking we investigate \textit{what} aspects of the overtaking process are critical to the dynamic interactions (section~\ref{sec:conceptual}) and \textit{how} to model these by assessing existing gap acceptance models (section~\ref{sec:review}). Finally, we provide a proof of concept of dynamic modeling of overtaking using the most suitable identified approach --- cognitive process modeling (section~\ref{sec:modeling}).

\section{Conceptual analysis of the overtaking interaction}\label{sec:conceptual}

To provide a road map for modeling human-AV interactions during overtaking, in this section we formulate requirements for the models based on the conceptual analysis of the overtaking maneuver.

When considering overtaking a single vehicle, three strategies can be applied: ‘piggy-backing’ (closely following another vehicle that overtakes the vehicle ahead), ‘flying’ (constant-speed overtaking) and ‘accelerating’ (slowing down behind the lead vehicle and then accelerate) \cite{Hegeman2005}. The scope of this study focuses on the latter as it is the commonly observed overtaking strategy \cite{Polus2000EvaluationOT}. 

The accelerating overtaking maneuver has been extensively analyzed by Hegeman et al. (\cite{Hegeman2005}). In our study, we use the twenty sub-tasks that follow from their analysis. To investigate these sub-tasks' interactive nature (explicit, implicit, or neither type of communication), we used the framework by Markkula et al. (\cite{Markkula2020}) by mapping the sub-tasks to interactive behaviors. Their proposed taxonomy describes seven non-mutually exclusive types of interactive behavior, of which three are related to moving in the traffic situation, another three to perceiving the traffic situation, and one to appreciating the traffic situation. In this framework, a distinction is made between \textit{implicit} and \textit{explicit} communication in interactions. Explicit communication (for example, the use of hand gestures or external human machine interfaces) only affects the other traffic participants’ movement and perception of the traffic situation. While implicit communication, such as making eye contact or accelerating to insist on the right of way, affects the movement and perception of both the ego vehicle and the other traffic participants. The nature of the interactions during overtaking provide a basis for the requirements for modeling of human-AV interactions. 

\begin{figure}[ht]
    \centering
    \includegraphics[width=0.45\textwidth]{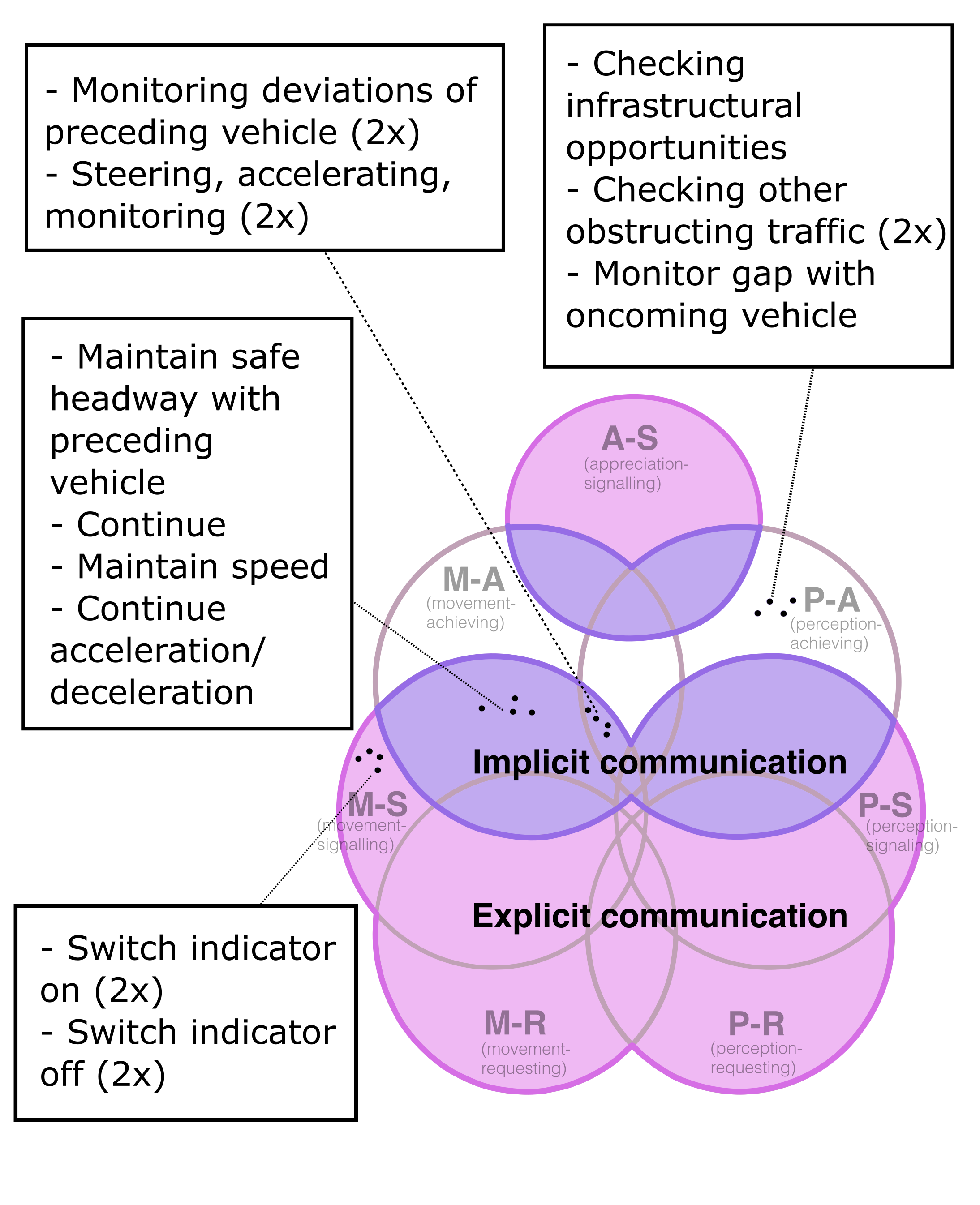}
    \caption{Mapping of the overtaking sub-tasks' \cite{Hegeman2005} interactive nature \cite{Markkula2020}. Sub-tasks with recurring instances are indicated within parentheses.}
    \label{fig:mapping}
\vspace{-5mm}
\end{figure}

Out of the 20 overtaking sub-tasks identified by Hegeman et al.~\cite{Hegeman2005}, we classified 16 as interactive (Figure \ref{fig:mapping}). Half of them relate to implicit communication, while the other half equally relates to either explicit or neither type of communication. Our finding that implicit communication is more present in the overtaking maneuver is in accordance with previous studies that show that explicit communication is rarely used by human road users \cite{Hegeman2005, Lee2021}. 

The effect of implicit communication such as motion dynamics on overtaking behavior plays a key role in gap acceptance. As an example, the speed of the oncoming vehicle significantly affects the critical gap (i.e. the gap where the probability to overtake or to stay is equal)~\cite{Farah2010}. Furthermore, to go from constant-speed scenarios to dynamic scenarios, dynamic interactions such as acceleration or deceleration by the oncoming vehicle should be considered as well. Here, we assume that these dynamics also affect the gap acceptance if perceptual thresholds are exceeded \cite{Mortimer, Schiff1990AccuracyOJ}. Human drivers exhibit adaptive behavior in response to changing road conditions and the behavior of other vehicles \cite{Rettenmaier2020}. Given that the overtaking decision process typically spans a duration of approximately 1 to 3 seconds on average~\cite{SEVENSTER2023329}, there might be dynamic changes in the environment within this time frame. As a result, incorporating models that capture response times can offer valuable insights into human behavior. Other key factors such as road geometry, driving style and driver’s demographics should be considered as well as they affect human overtaking behavior \cite{Farah2010, Llorca2013, Vlahogianni2013}.

To summarize, the requirements for modeling gap acceptance in human-AV overtaking interaction are:

\begin{enumerate}
    \item Describing \textit{dynamic} interactions (i.e. taking into account AV’s motion dynamics)
    \item Describing human's response time (the moment of accepting gap)
    \item Possibility of incorporating other key factors affecting gap acceptance (i.e. driver’s demographic characteristics, driving style, and road geometry)
\end{enumerate}

\section{Assessment of gap acceptance models}\label{sec:review}

As we argued in the previous section, gap acceptance is a key element of the overtaking process, and therefore human-AV interaction models should incorporate it. Existing models of gap acceptance (not only in overtaking but also several other traffic scenarios such as entering intersections \cite{Pollatschek}, lane-changing \cite{Toledo2003ModelingIL}, and pedestrian crossing \cite{Papadimitriou2012}) can be based on: logistic regression \cite{Farah2010, Yannis2013}, machine learning \cite{Nagalla2017, Kadali}, algorithmic modeling \cite{Ottomanelli2010, Stefansson2020}, agent-based modeling
\cite{Kang2017, Zohdy2012} and cognitive modeling \cite{Pekkanen2022, ZgonnikovAbbinkMarkkula}.

We assessed different classes of gap acceptance models according to the three criteria mentioned in the previous section (Table \ref{tab:assessment}). Following the approaches listed in Table \ref{tab:assessment}, we found that two approaches are promising in modeling human decision making during overtaking interactions with AVs. These are agent-based/game-theoretic models and cognitive models. 

Agent-based models describe interactions between traffic participants as agents in a game-theoretic setting (e.g.~\cite{Kang2017,Zohdy2012}). They are able to handle dynamic changes of vehicle dynamics during the interactions. However, implementing response time in game-theoretic models has so far only been done indirectly by using receding horizon control \cite{Meng}. Then to also include additional factors these need to be combined with extended agent-based models \cite{Zohdy2012}. Furthermore, game-theoretic assumptions of agents having perfect knowledge of each other and not needing to communicate may not hold in real-world scenarios~\cite{siebinga2023model}. Game-theoretic models also rely on assumptions about payoff functions that may not reflect human decision making. 

Cognitive models of gap acceptance describe cognitive processes that underlie human decision making in traffic, building up on fundamental research in cognitive psychology and neuroscience~\cite{Ratcliff1978ATO,RATCLIFF2016260}. One class of cognitive models that is becoming increasingly popular in modeling traffic interactions is drift-diffusion models (DDMs, e.g.~\cite{Pekkanen2022,ZgonnikovAbbinkMarkkula}). DDMs naturally capture response times~\cite{Ratcliff1978ATO} and can incorporate changes in vehicle dynamics during the interaction~\cite{Pekkanen2022,ZgonnikovAbbinkMarkkula}. Furthermore, in comparison to agent-based models, the DDM framework provides a simpler approach to incorporating the factors affecting gap acceptance (age and gender, and road geometry) by adjusting model parameters~\cite{Theisen}. Challenges include representing these factors simultaneously and incorporating them in advanced DDM models that are needed to model dynamic interactions. Despite these challenges, we conclude that cognitive models and the DDM in particular hold promise for more realistic models of overtaking in human-AV interactions.

\begin{table}[ht]
    \centering
        \caption{Assessment of gap acceptance models}
    \begin{tabular}{l|l|l|l}
        \textbf{Model type} & \textit{Dynamic} & \textit{Response time} & \textit{Influencing factors}  \\ \hline
        Algorithmic & no & no & no \\
        Agent-based  & yes & indirect & indirect \\
        Cognitive  & yes & yes & indirect  \\
        Logistic  & no & no & yes \\
        Machine learning  & indirect & no & yes \\             
    \end{tabular}

    \label{tab:assessment}
\end{table}

Earlier work showed that the DDM holds promise in accurately predicting gap acceptance in left-turn decisions \cite{ZgonnikovAbbinkMarkkula, nudge}.  However, in contrast to the left-turn decision that are made at low speeds, in overtaking the human driver is already at a relatively high speed when initiating the decision-making process. Furthermore, in overtaking multiple sources of dynamic evidence may affect gap acceptance, considering both the presence of the lead vehicle and the oncoming AV. DDMs used in other traffic scenarios such as unprotected left-turns \cite{ZgonnikovAbbinkMarkkula, nudge} and pedestrian crossing \cite{Pekkanen2022} therefore may need to be adapted to accommodate the ego vehicle's initial speed and the lead vehicle’s presence. 

\section{Modeling human decision making in overtaking: A proof-of-concept}\label{sec:modeling}
To investigate the feasibility of the cognitive modeling approach for overtaking, here we test several version of the drift-diffusion model using the data on human overtaking decisions previously collected in a driving simulator~\cite{SEVENSTER2023329}. The model fitting and simulation code used in this case study is available \href{https://github.com/shamohammad/Overtaking_DDM}{online}.

\subsection{Dataset}
\label{sec:data}
A prerequisite of cognitive process modeling using the drift-diffusion models is measuring the response time. Sevenster et al.~\cite{SEVENSTER2023329} offered a simple way of measuring accepted and rejected response times in overtaking, and explored the effect of two situation-specific factors (distance gap and ego-vehicle velocity) on the response times measured in a driving simulator experiment. 
The measures of Sevenster et al. (\cite{SEVENSTER2023329}) included 2097 overtaking decisions collected from 25 participants, with varying initial gap to the oncoming vehicle (160 or 220 meters) and the initial ego-vehicle velocity as a free variable. It included the decision outcome and the corresponding response time as the dependent variables. To be able to model this dataset, we filtered it by removing any measures with unrealistic response times, missing values, and null values. The remaining data (N=1758) was used for further analysis.

The continuous nature of the free initial ego-vehicle velocity variable impedes model fitting using existing fitting tools such as \textit{pyddm} \cite{shinn}. Therefore, in this study, this variable has been clustered into three initial velocities, and by this transforming the problem to a 2x3 factorial design (2 initial distance conditions, 3 initial velocity conditions). We have opted to exclude measures relating to the lead vehicle such as following distance, since clustering these as well would significantly reduce the amount of data for each set of conditions. 

Based on their data, Sevenster et al.~\cite{SEVENSTER2023329} highlighted the following relationships between the initial setup of the overtaking scenario and the resulting human behavior and response times:
\begin{itemize}
    \item Probability of accepting the gap increases with initial distance to the oncoming vehicle.
    \item Probability of accepting the gap increases with initial velocity of the ego vehicle.
    \item Response times in rejected gaps are on average higher than in accepted gaps. 
    \item Response times in both accepted in rejected gaps increase with initial distance.
    \item Response times in accepted gaps decrease with initial velocity.
    \item Response times of rejected gaps remain constant regardless of the initial velocity.
\end{itemize}

In what follows, we evaluate how well different candidate cognitive models can capture human behavior according to these findings. 

\subsection{Cognitive modeling}
\subsubsection{Basic drift-diffusion model and its applications to traffic}
\begin{figure}[ht]
    \centering
    \includegraphics[width=\columnwidth]{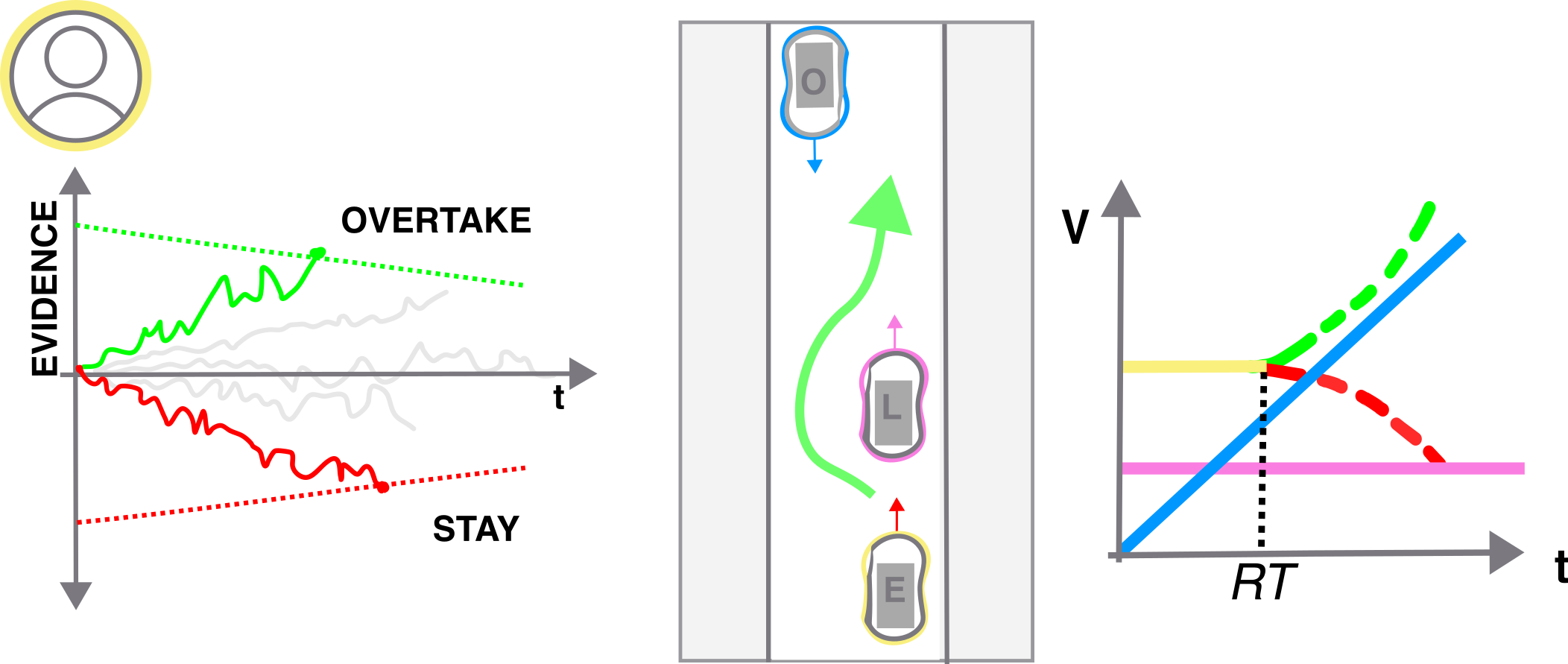}
    \caption{Visualization of gap acceptance decision making in overtaking. Depending on the gap to the oncoming vehicle (blue), the human driver of the ego vehicle (yellow) can decided either to reject the gap and stay in the lane (red trajectory) or to accept the gap (green trajectory) and overtake the slow lead vehicle. According to the drift-diffusion model, this decision can be represented as bounded accumulation of noisy evidence over time.}
    \label{scenario}
\end{figure}

We employed the drift-diffusion modeling framework~\cite{Ratcliff1978ATO} to explain participants' behavior and response times in our experiment. This framework is based on evidence accumulation, where humans integrate relevant perceptual information over time (Figure \ref{scenario}). Accumulation is a noisy process that continues until the evidence in favor of one alternative reaches a predetermined boundary. Despite its simplicity, DDMs have been successful in explaining various behavioral effects of decision context on outcomes and response times \cite{RATCLIFF2016260}.

Mathematically, the drift-diffusion model represents the choice between two options as a random process, where evidence $x$ accumulates based on a drift rate $s(t)$ (momentary evidence favoring one option over the other) and diffusion (random noise $\epsilon(t)$): 

\begin{equation}
\frac{dx}{dt}=s(t) + \epsilon(t).
\label{eq:ddm}
\end{equation}

Accumulation stops when the accumulated evidence crosses an upper $x = b(t)$ or lower decision boundary $x = -b(t)$. 

Recent applications of DDM to gap acceptance~\cite{Pekkanen2022, ZgonnikovAbbinkMarkkula} consider the drift rate $s(t)$ to capture dynamically changing gap sizes and time-varying decision boundaries $b(t)$ to reflect choice urgency. Such models were able to capture decision outcomes and response times of human decision makers. However, they cannot be directly used for our overtaking scenario because they do not incorporate the initial velocity that the human driver has at the start of the decision. As previous studies have shown, this velocity affects the decision and therefore it needs to be incorporated in one of the components of the DDM.

\subsubsection{Drift-diffusion model of overtaking}
Here, we build upon the previously proposed left-turn gap acceptance model \cite{ZgonnikovAbbinkMarkkula} by incorporating the initial velocity of the ego vehicle in the different components of the model (drift rate, decision boundary, initial decision bias). We then investigate which of the resulting 8 versions of the model better describes the data of Sevenster et al.~\cite{SEVENSTER2023329}. 

Each of the tested models includes four main components. First, the drift rate $s(t)$ is a function of time-to-arrival ($TTA$) and distance $d$ between the ego vehicle and the oncoming vehicle and possibly the initial velocity of the ego vehicle $v_0$

\begin{equation}
 s(t) = \alpha(TTA(t) + \beta d(t) - \theta_s) \label{eq:drift}   
\end{equation}
 \begin{equation}
s(t) = \alpha(TTA(t) + \beta d(t) + \gamma v_0  - \theta_s) \label{eq:drift_extended},
\end{equation}
where $\alpha>0$, $\beta>0$, $\gamma>0$ and $\theta_s>0$ are free parameters. We define $x$ as a measure of \textit{relative} evidence, with positive values indicating support for the ``Overtake'' decision and negative values favoring the ``Stay'' decision at a given moment $t$. Intuitively, as the gap between the decision maker and the oncoming vehicle (a combination of $d$ and $TTA$) increases (e.g., when the opposing vehicle decelerates) relative to a critical value $\theta_s$, the drift rate becomes more positive. This implies a higher likelihood of the decision maker leaning towards the Overtake decision. Conversely, they are more likely to arrive to the Stay decision when the drift rate becomes more negative. As the initial speed of the ego vehicle positively affects the probability of accepting the gap~\cite{SEVENSTER2023329}, these effects are amplified when including the initial velocity in the drift rate. 

Second, the decision boundary collapses with either $TTA(t)$, or with all the kinematic variables affecting the drift rate $s(t)$. 
\begin{equation}
 b(t) = \pm \frac{b_0}{1+e^{-k(TTA(t)-\tau)}} \label{eq:bound}    
\end{equation}
\begin{equation}
b(t) = \pm \frac{b_0}{1+e^{-k(TTA(t) + \beta d(t) -\theta_s)}} 
\label{eq:bound_drift}
\end{equation}
\begin{equation}
 b(t) = \pm \frac{b_0}{1+e^{-k(TTA(t) + \beta d(t) + \gamma v_0 -\theta_s)}}.
 \label{eq:bound_drift_extended}   
\end{equation}
Intuitively, with lower values of $TTA$ and $d$ the decision maker experiences stronger urgency to make the decision, which is reflected by boundary $b(t)$ decreasing with the gap size (similar to~\cite{ZgonnikovAbbinkMarkkula}).

Third, the initial bias $Z$ defines the starting position of the evidence accumulation process (i.e. $x(t_0)=Z$)
\begin{equation}
Z = C_z 
\label{eq:bias_constant}    
\end{equation}
\begin{equation}
 Z = \frac{2b(t_0)}{1+e^{-b_z(v_0-\theta_z)}} - b(t_0) \label{eq:bias_vel},   
\end{equation}
where a value of $Z < 0$ indicates an initial bias towards the Stay decision, while $Z > 0$ indicates a bias towards the Overtake decision. This bias can be represented by a constant value $C_z$ (Eq. \eqref{eq:bias_constant}) or can vary based on the initial velocity $v_0$ (Eq. \eqref{eq:bias_vel}). In the latter case, relatively higher and lower initial speeds correspond to a bias towards the Overtake and Stay decision, respectively.

Fourth, for all models the non-decision time (the duration of the cognitive processes unrelated to decision-making, such as perceptual and motor delays) is assumed to follow the normal distribution
\begin{equation}
 t^{ND} \in \mathcal{N}(\mu_{ND},\sigma_{ND}),\quad \mu_{ND}>0,\, \sigma_{ND}>0.   
 \label{eq:ndt}
\end{equation}

The eight model variants resulting from different combinations of the model components are shown in Table \ref{tab:modelvariations}. The odd-numbered models use a constant bias, while the even-numbered models use a bias depending on the initial speed. Models 1, 2, 5 and 6 have their drift rate depending on the $TTA(t)$ and $d(t)$, whereas Models 3, 4, 7 and 8 also include the initial speed. The decision boundaries of Models 1 to 4 decrease with the $TTA$, while Models 5 to 8 use decision boundaries depending on all kinematic variables affecting their respective drift rate function. The simplest model (M6) contains 8 free parameters ($\alpha$, $\beta$, $\theta_s$, $b_0$, $k$, $Z$, $\mu_{ND}$, $\sigma_{ND}$) and the most extensive model (M4) contains 11 free parameters ($\alpha$, $\beta$, $\gamma$, $\theta_s$, $b_0$, $k$, $\tau$, $\theta_z$, $b_z$, $\mu_{ND}$, $\sigma_{ND}$).

\begin{table*}[ht]
    \centering
    \caption{Tested variations of the generalized drift-diffusion model~\eqref{eq:ddm}. The number of parameters in the last column includes the two non-decision time parameters $\mu_{ND}$, $\sigma_{ND}$.}
\begin{tabular}{lclclclc}
Model & Drift rate $s(t)$ & Eq. & Decision boundary $b(t)$ & Eq. & Initial bias $ -b(t_0) < Z < b(t_0)$ & Eq. & \# parameters  \\
M1 & $\alpha(TTA(t) + \beta d(t) - \theta_s)$ & \eqref{eq:drift} &$ \pm \frac{b_0}{1+e^{-k(TTA(t)-\tau)}}$ &\eqref{eq:bound} & $C_z$ & \eqref{eq:bias_constant} & 9\\
M2 & $\alpha(TTA(t) + \beta d(t) - \theta_s)$ &\eqref{eq:drift} &$ \pm \frac{b_0}{1+e^{-k(TTA(t)-\tau)}}$& \eqref{eq:bound} & $\frac{2b(t_0)}{1+e^{-b_z(v_0-\theta_z)}} - b(t_0) $& \eqref{eq:bias_vel} & 10\\
M3 & $\alpha(TTA(t) + \beta d(t) + \gamma v_0 - \theta_s)$& \eqref{eq:drift_extended} &$ \pm \frac{b_0}{1+e^{-k(TTA(t)-\tau)}}$ &\eqref{eq:bound} & $C_z$& \eqref{eq:bias_constant} & 10\\
M4 & $\alpha(TTA(t) + \beta d(t) + \gamma v_0 - \theta_s)$& \eqref{eq:drift_extended}  &$ \pm \frac{b_0}{1+e^{-k(TTA(t)-\tau)}}$& \eqref{eq:bound}& $\frac{2b(t_0)}{1+e^{-b_z(v_0-\theta_z)}} - b(t_0) $& \eqref{eq:bias_vel} & 11 \\
M5 & $\alpha(TTA(t) + \beta d(t) - \theta_s)$& \eqref{eq:drift} &$ \pm \frac{b_0}{1+e^{-k(TTA(t) + \beta d(t) - \theta_s)}}$ &\eqref{eq:bound_drift} & $C_z$& \eqref{eq:bias_constant} & 8\\
M6 & $\alpha(TTA(t) + \beta d(t) - \theta_s)$& \eqref{eq:drift} &$ \pm \frac{b_0}{1+e^{-k(TTA(t) + \beta d(t) - \theta_s)}}$ &\eqref{eq:bound_drift}  & $\frac{2b(t_0)}{1+e^{-b_z(v_0-\theta_z)}} - b(t_0)$& \eqref{eq:bias_vel} & 8 \\
M7 & $\alpha(TTA(t) + \beta d(t) + \gamma v_0 - \theta_s)$& \eqref{eq:drift_extended}  &$ \pm \frac{b_0}{1+e^{-k(TTA(t) + \beta d(t) + \gamma v_0 - \theta_s)}}$& \eqref{eq:bound_drift_extended}& $C_z$& \eqref{eq:bias_constant} & 9\\
M8 & $\alpha(TTA(t) + \beta d(t) + \gamma v_0 - \theta_s)$ & \eqref{eq:drift_extended}  &$ \pm \frac{b_0}{1+e^{-k(TTA(t) + \beta d(t) + \gamma v_0 - \theta_s)}}$& \eqref{eq:bound_drift_extended} & $\frac{2b(t_0)}{1+e^{-b_z(v_0-\theta_z)}} - b(t_0)$& \eqref{eq:bias_vel} & 10\\
\end{tabular}
\label{tab:modelvariations}
\end{table*}

\subsubsection{Model fitting and evaluation}
Our goal was to examine whether extended models could depict the behavior of the "average" participant in the dataset. Although it is possible to fit the model to each participant's data individually, providing insights into individual differences (see e.g.~\cite{ZgonnikovAbbinkMarkkula}), it requires a separate investigation beyond the scope of this study. Instead, we evaluated the models' qualitative match to the data reported in~\cite{SEVENSTER2023329} according to the observations listed in the end of Section~\ref{sec:data}. 

The fitting of the models involved utilizing the differential evolution optimization technique and Bayesian information criterion, as implemented in the \textit{pyddm} framework, a Python package specifically designed for DDM fitting \cite{shinn}.

\subsubsection{Comparing models and data}

\begin{figure*}[ht]
    \centering
    \begin{subfigure}{0.32\textwidth}
    \includegraphics[width=\linewidth]{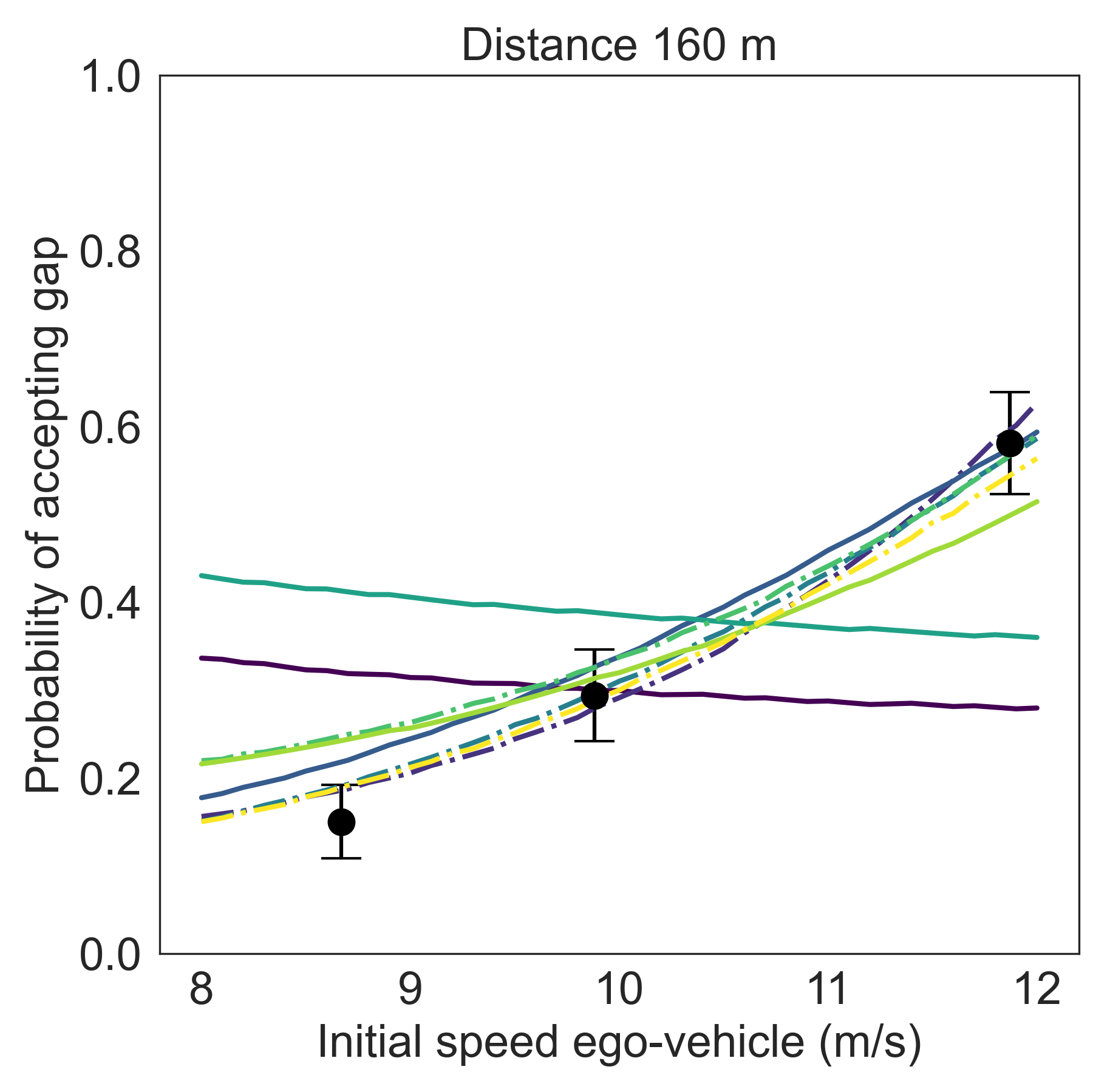}
    \captionsetup{justification=centering}
    \end{subfigure}
    \hfill
    \begin{subfigure}{0.32\textwidth}
    \includegraphics[width=\linewidth]{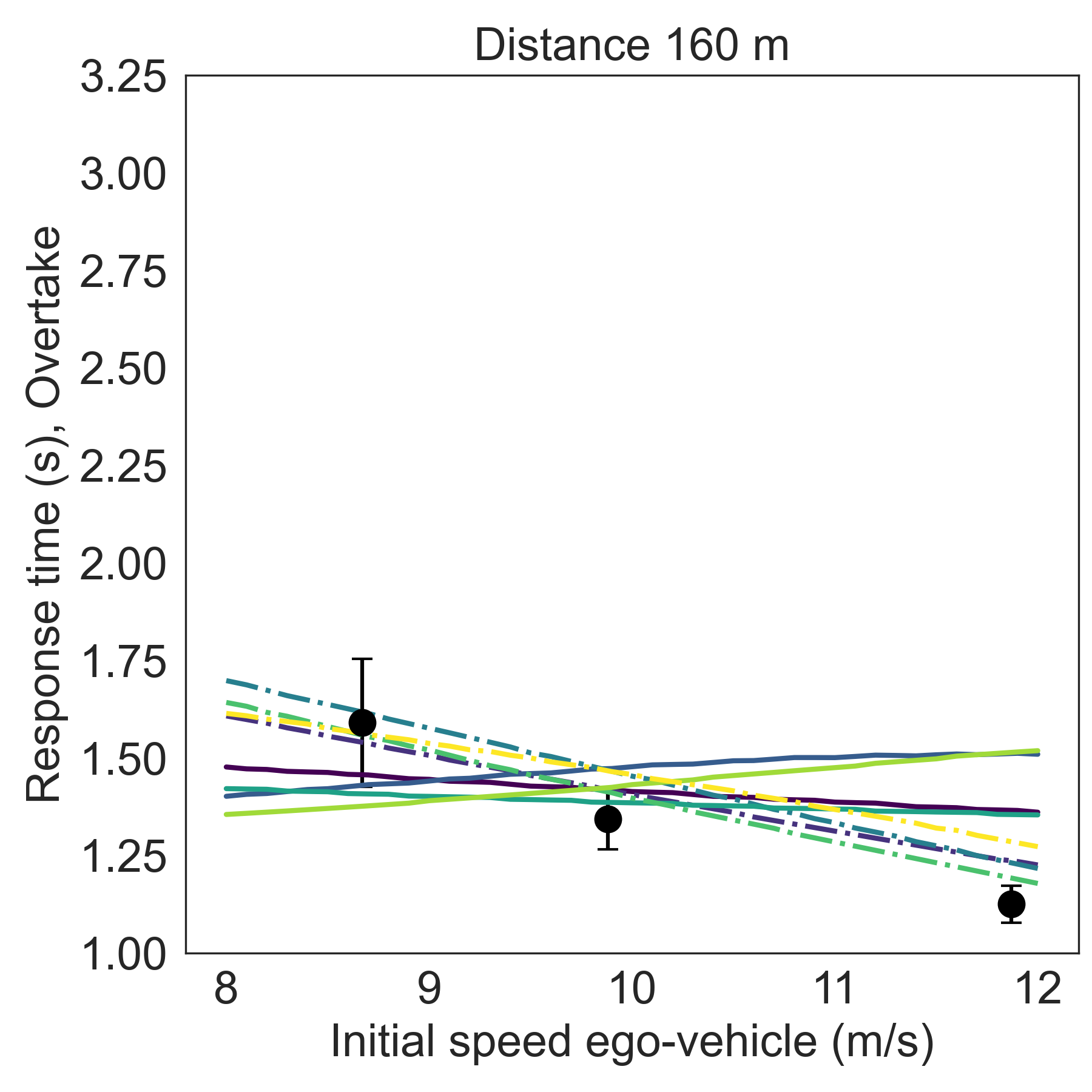}
    \end{subfigure}
    \hfill
        \begin{subfigure}{0.32\textwidth}
    \includegraphics[width=\linewidth]{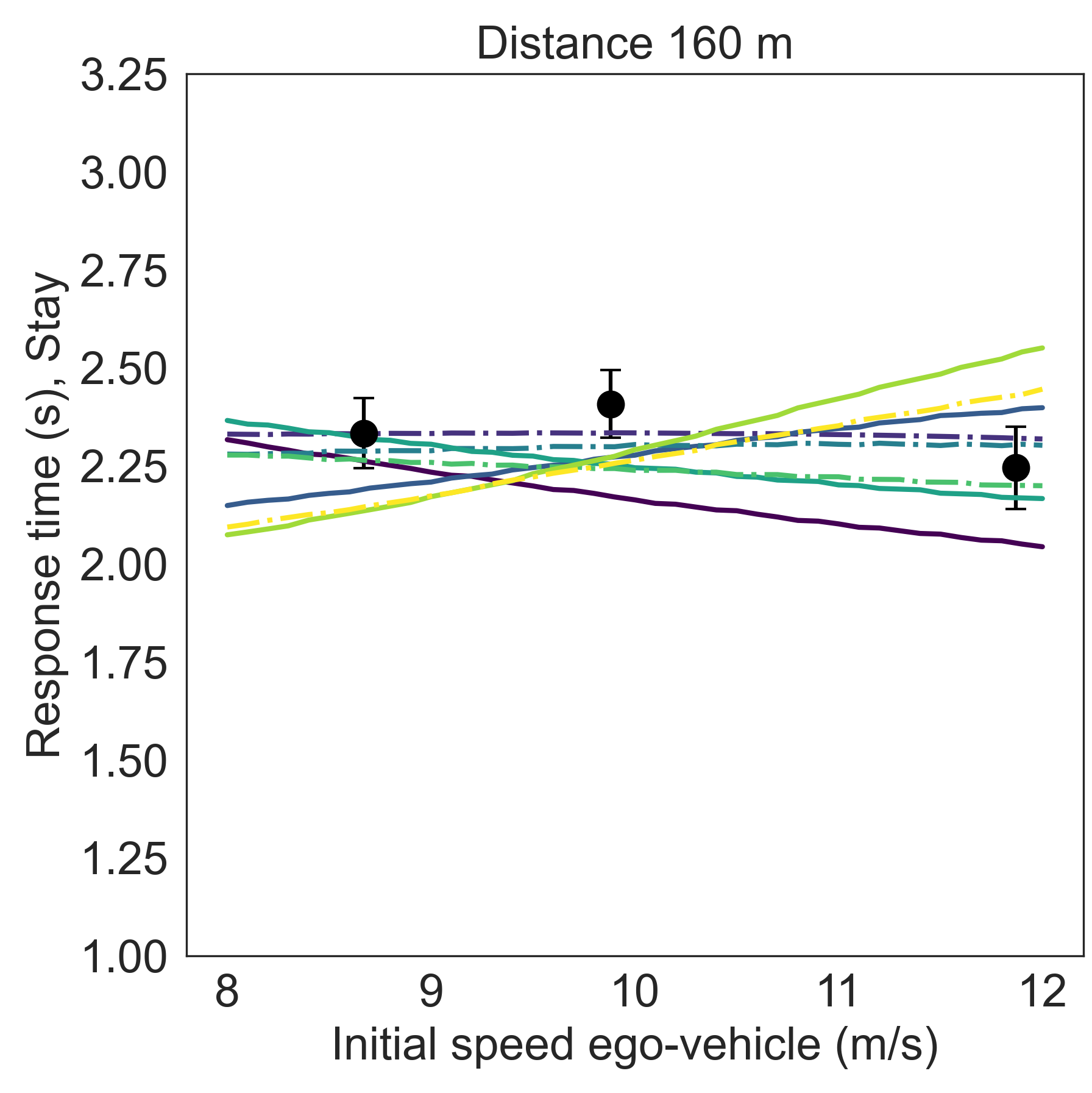}
    \end{subfigure}
    \\
    \begin{subfigure}{0.32\textwidth}
    \includegraphics[width=\linewidth]{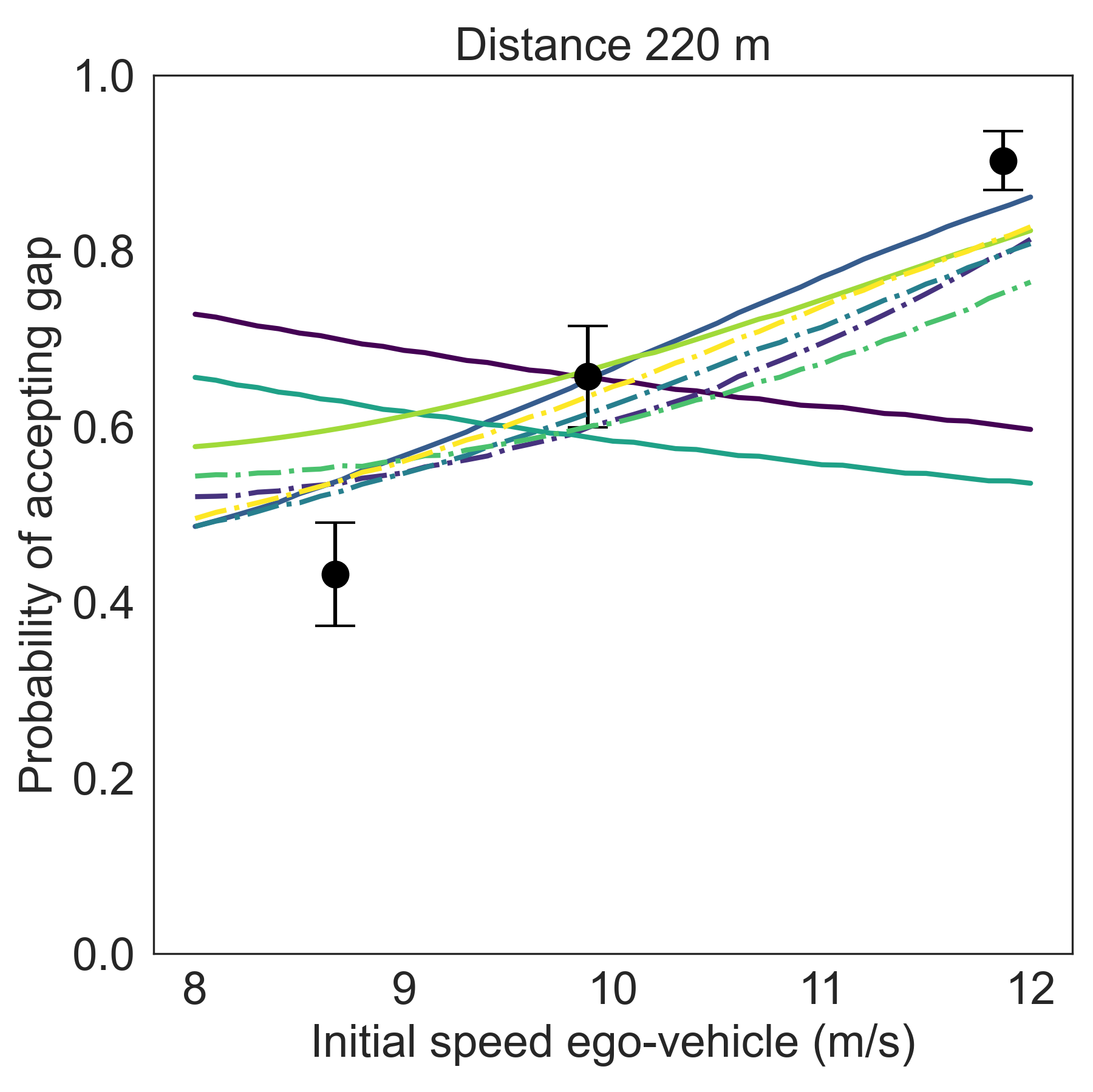}
    \end{subfigure}
    \hfill
    \begin{subfigure}{0.32\textwidth}
    \includegraphics[width=\linewidth]{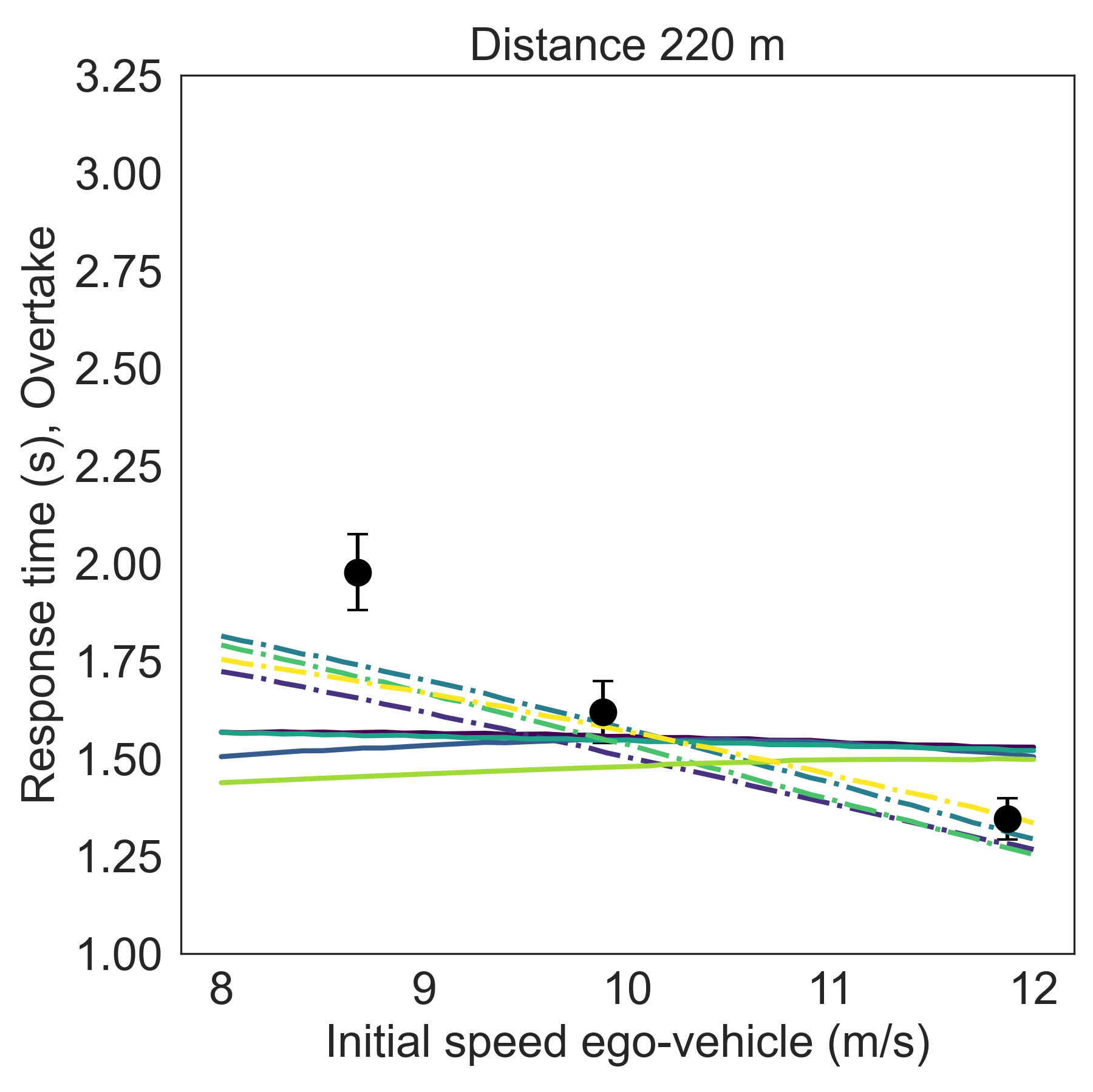}
    \end{subfigure}
    \hfill
    \begin{subfigure}{0.32\textwidth}
    \includegraphics[width=\linewidth]{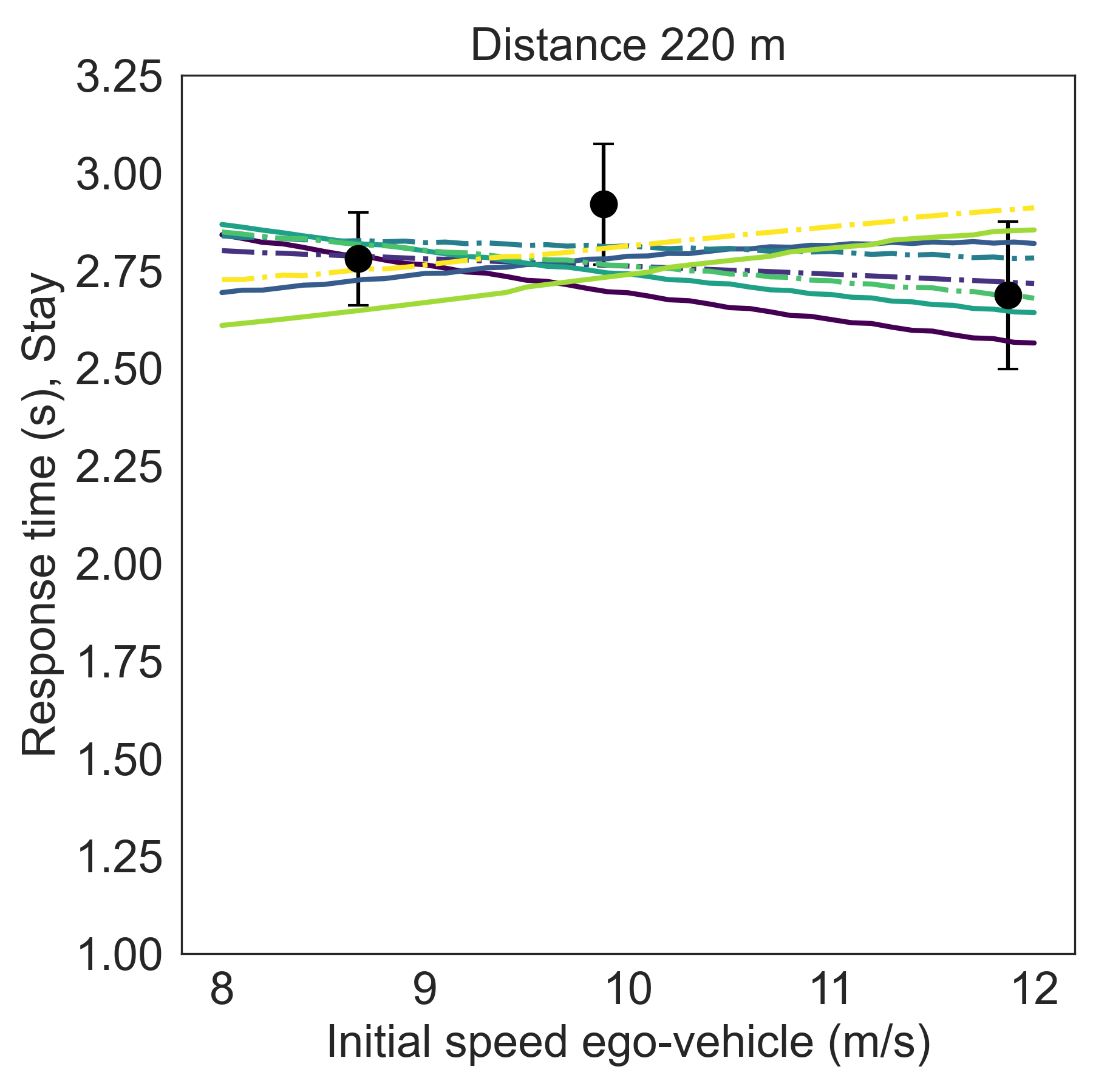}
    \end{subfigure}
    \\
    \begin{subfigure}{\textwidth}
    \includegraphics[width=\linewidth]{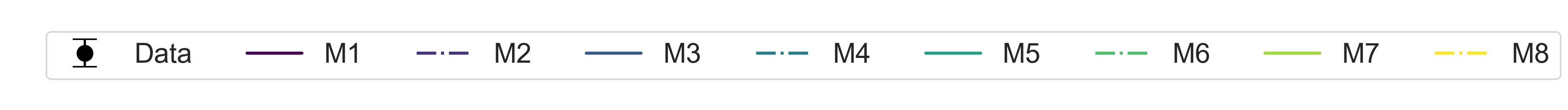}
    \end{subfigure}
    \caption{Simulated model results compared to the data of Sevenster et al.~\cite{SEVENSTER2023329}. The error bars represent the standard error of the mean.}
    \label{fig:results}
\end{figure*}

\begin{table*}[htbp]
    \centering
    \caption{Assessment of candidate drift-diffusion models according to the experimental findings of Sevenster et al.~\cite{SEVENSTER2023329}.}
    \begin{tabular}{l|cccccccc}
        \multicolumn{1}{c|}{Finding} & M1 & M2 & M3 & M4 & M5 & M6 & M7 & M8 \\
        \hline
        Probability of accepting the gap increases with initial distance to the oncoming vehicle. & \Checkmark & \Checkmark  & \Checkmark & \Checkmark & \Checkmark  & \Checkmark & \Checkmark   & \Checkmark \\
        Probability of accepting the gap increases with initial velocity of the ego vehicle. & \textbf{X} & \Checkmark & \Checkmark &\Checkmark & \textbf{X}  & \Checkmark & \Checkmark & \Checkmark  \\
        Response times in rejected gaps are on average higher than of accepted gaps. & \Checkmark & \Checkmark  & \Checkmark  & \Checkmark  & \Checkmark & \Checkmark  & \Checkmark  & \Checkmark  \\  
        Response time in both accepted in rejected gaps increases with initial distance. & \Checkmark & \Checkmark & \textbf{X} & \Checkmark & \Checkmark & \Checkmark & \textbf{X} &  \Checkmark \\
        Response times in accepted gaps decrease with initial velocity. & \textbf{X} & \Checkmark & \textbf{X} & \Checkmark & \textbf{X} & \Checkmark & \textbf{X} & \Checkmark \\
        Response times of rejected gaps remain constant regardless of initial velocity. & \textbf{X} & \Checkmark & \textbf{X} & \Checkmark & \textbf{X} & \Checkmark & \textbf{X} & \textbf{X} \\
        \hline
        \textbf{Total} & \textbf{3/6} & \textbf{6/6} & \textbf{3/6} & \textbf{6/6} & \textbf{3/6} & \textbf{6/6} & \textbf{3/6} & \textbf{5/6}
    \end{tabular}
    \label{tab:model_assessments}
\end{table*}

We found that the eight tested models differed substantially in regards to their qualitative match with the observed human behavior (Figure~\ref{fig:results}, Table~\ref{tab:model_assessments}). 

The models that did not include the ego vehicle's initial speed $v_0$ in any of the components (M1 and M5) predictably could not capture the increase of probability of accepting the gap with $v_0$. The other six models could all account for probability of accepting the gap, making it essential to consider response time as the measure that can help distinguish between candidate models further.

For response times, the results differ considerably between odd- and even-numbered models (Table \ref{tab:model_assessments}). The odd-numbered models, i.e. models with a constant initial bias, struggle to consistently describe the effect of initial velocity on response times (in both accepted and rejected gaps). On the other hand, among the models that do include velocity-dependent initial bias, M8 captures 5 out of 6 qualitative patterns, and M2, M4 and M6 even describe them all.

The most successful models, M2, M4 and M6, contain respectively 10, 11 and 9 free parameters. The differences between these three models can be found in the decision boundary: decision boundaries of M2 and M4 collapse only with $TTA(t)$, while M6's boundary collapses with $TTA(t)$ and $d(t)$. Furthermore, in contrast to the drift rate used in M4, M2 and M6 do not have the initial velocity included in theirs. Lastly, M6 reuses parameters of the drift rate in the boundary function, therefore consolidating the total amount of free parameters. Therefore, we conclude that M6 is the simplest model that can describe all qualitative patterns previously observed in human behavior. This model hypothesizes drift rate and decision boundary that both depend on the same linear combination of TTA and distance, and the decision bias that scales with the initial velocity of the ego vehicle. The resulting fitted model parameters for M6 were $\alpha = 0.07$, $\beta = 0.11$, $\theta_s = 47$, $b_0 = 2.8$, $k = 0.02$, $b_z = 0.14 $, $\theta_z = 5.8$, $\mu_{ND} = 1.0 $, $\sigma_{ND} = 0.27$.

\section{Discussion}
Human decision-making in traffic involves high stakes, especially during overtaking where there is an increased risk of a head-on collision between two vehicles at high speed. Understanding and predicting human overtaking behavior can lead to safer interactions  on the road. This paper makes a step towards such understanding by conceptually analyzing the overtaking process and applying the cognitive modeling approach to describe the dynamic decision-making process of human drivers in overtaking.  

Our study highlights that drift-diffusion model can be transferred to complex traffic scenarios, such as the overtaking maneuver. This represents a major step forward compared to simpler traffic scenarios that have been modelled with drift-diffusion models before~\cite{nudge, Pekkanen2022,ZgonnikovAbbinkMarkkula}. Potentially, decision making in other dynamic interactive maneuvers, such as  merging from on-ramps and lane changing on highways, can be described as well by these models. 

An important limitation of this study however is that our conceptual analysis only mapped the interactive behaviors of the ego vehicle. To fully conceptualize interactions in any traffic scenario, all traffic participants should be taken into account \cite{Markkula2020}. Furthermore, the role of the lead vehicle has not been explored thoroughly when modeling overtaking behavior even though empirical studies show that the lead vehicle's dynamics affect gap acceptance \cite{Farah2010}. Lastly, the effect of driving characteristics, such as age and gender, on gap acceptance is not investigated in our proof-of-concept study due to the lack of existing datasets that measure response times in large samples of participants. Previous studies highlighted that such characteristics affect gap acceptance in overtaking~\cite{Llorca2013}, so addressing them in DDMs could be useful when modeling participant-specific overtaking behavior. 

This paper goes beyond existing constant-speed gap acceptance studies in overtaking by providing DDM components that can potentially describe overtaking behavior when interacting with an oncoming vehicle that changes its dynamics during the overtaking maneuver. Given the characteristic response times (1 to 3 seconds on average~\cite{SEVENSTER2023329}), such dynamic changes can affect the ongoing gap acceptance decision. This represents an important potential point of influence for AVs to manage the interaction with the human-driven vehicles~\cite{Miller2022a,nudge,Rettenmaier2021}. Future research should therefore examine how such dynamic interactions with an oncoming AV can be studied empirically and modelled. 

Our work has potential practical applications for safer human-AV interactions in overtaking. Cognitive models like the DDM can be used to enhance training and validation of existing interactive-aware controllers \cite{Siebinga} in the case only limited training and validation data are available \cite{Trafton2020}. Furthermore, models like DDM could be used for better predictions in human-AV interactions, which can benefit traffic safety. Firstly, the risk of head-on collisions can be reduced by AVs anticipation of overtaking behavior of other road users \cite{Sadigh2016PlanningFA}. Secondly, traffic flow can also become more efficient through trajectory planning of AVs \cite{Rettenmaier2020}. Further research is needed however on utilizing the potential of DDMs for behavior prediction in gap acceptance~\cite{Schumann}.

\section{Conclusion}
This study shows the promise of using drift-diffusion models, a subset of cognitive process models, to predict human gap acceptance in overtaking. Our results can be used in future research to predict human overtaking behavior when dynamically interacting with an oncoming AV. We believe that this will help to understand how AVs could control their interaction strategy to contribute to safer and more efficient traffic. More generally, this study exemplifies how simple cognitive process models can help us to understand and possibly improve human-AV interactions in complex traffic scenarios.

\addtolength{\textheight}{-1cm}   







\bibliographystyle{jabbrv_ieeetr}
\bibliography{IEEEabrv,main.bbl}
\end{document}